\documentstyle[color,preprint,aps,pre,eqsecnum,psfig,amssymb]{revtex}

\begin{document}


\title{\bf Roll and square convection in binary  liquids: a few--mode Galerkin model}

\author{S.~Weggler, B.~Huke, and M.~L\"ucke}
\address{Institut f\"ur Theoretische Physik, Universit\"at des Saarlandes,
Postfach 151150, D--66041 Saarbr\"ucken, Germany}

\date{\today}

\maketitle

\begin{abstract}
We present a few--mode Galerkin model for convection in binary fluid layers
subject to an approximation to realistic horizontal boundary conditions at positive separation
ratios. The model exhibits convection patterns in form of rolls and squares. 
The stable squares at onset develop into stable rolls at higher thermal driving.
In between, a regime of a so-called crossroll structure is found.
The results of our few--mode model are in good agreement with both experiments and
numerical multi--mode simulations.
\end{abstract}

\pacs{47.20.Bp 47.54.-r 47.20.Ky}

\tightenlines

\newcommand{\Nabla}{\mbox{\bf\boldmath $\nabla$}}
\renewcommand{\vec}[1]{{\bf #1}}
\renewcommand{\u}[1]{\underline{#1}}
\newcommand{\Overrightarrow}[1]{\stackrel{\textstyle\rightarrow}{#1}}

\section{Introduction}
\label{sec:I}
Compared to convection in ordinary one--component 
fluids the spatiotemporal properties of binary fluids are far more 
complex. The evolution of the concentration field is governed by the 
interplay of typically strong nonlinear convective transport and mixing,
weak dissipative solutal diffusion, and the Soret effect 
\cite{LL66,PL84}. The Soret effect generates concentration gradients in response to the 
externally applied temperature difference and to local temperature 
gradients. The strength of the Soret coupling is measured by the 
dimensionless separation ratio $\psi$. The driving
mechanisms are therefore controlled by the Rayleigh number $R$,
measuring the temperature stress, and by the separation ratio, i.e. the
solutal driving.

In the present paper we focus on 2--dimensional (2D) convective structures 
consisting of straight parallel rolls in one lateral direction, and 3--dimensional (3D) 
structures that look like a nonlinear superposition of two perpendicular 
roll sets. These structures exist at positive separation ratios $\psi>0$ and arise 
from a stationary supercritical bifurcation either directly out of the ground state 
or out of a primary convective state.

At onset, the convection is driven mainly by the solutal gradient established via 
the Soret effect. Therefore this regime is called Soret regime in the literature, 
see e.~g.\ \cite{ML88}. The stable convection structure is typically a 3D pattern 
with square symmetry.

For larger Rayleigh numbers the concentration homogenizes and the fluid behaves 
more like a pure fluid. Convection is now driven mainly by
temperature gradients. The stable convection pattern is a 2D roll pattern.

In the intermediate regime, where the stability changes from stationary squares
to stationary rolls, there exists another 3D structure, the so--called crossroll 
pattern. This structure
bifurcates out of the stable square branch and merges with the roll branch at
higher $R$. Finally, for slow solutal diffusion, the competition between 
rolls and squares leads to an oscillating behavior in an interval of heating
rates around the bifurcation point from squares to crossrolls.

The bifurcation scenario described above has been verified by several
experimental groups, e.~g.~\cite{MS86,GPC85}. A detailed theoretical 
insight into the bifurcation scenario has been provided by numerical simulations of 
Ch.~Jung et al.\ \cite{JHL98}.

The numerical analysis \cite{ML88,JHL98,HLM98,HLBJ00}
elucidating the influence of the spatiotemporal behavior of the concentration
field on various properties of convective states for negative and positive
separation ratio has clearly shown that
the success of a model description sensitively hinges upon the representation
of the concentration field. The representation has to
capture the essence of the spatiotemporal structures
following  from the combined action of strong nonlinear advection and weak
diffusion on one hand and the generation of Soret induced concentration
currents by temperature gradients on the other hand.

For $\psi>0$ a model with few degrees of freedom that
reproduces all essentials of the bifurcation
behavior of the flow amplitude is presently not
available. The first attempt to model this bifurcation topology by M\"uller et
al. could only generate stable rolls and unstable squares \cite{ML88}.

Our paper aims at filling this gap. We present a few--mode
Galerkin model which is based upon a careful analysis of
the concentration balance \cite{ML88,HLM98,HLBJ00,H96} in liquid mixtures. With it we
explain the whole bifurcation scenario
from stable squares at onset up to stable rolls.
The model is an extension of the few--mode model, which was presented in
\cite{HLM98} for negative separation ratios.

We introduce the system and formulate the theoretical task in
Sec.~\ref{sec:II}\,. The main body of this paper consists of the two following
sections: In Sec.~\ref{sec:III} we construct the Galerkin model and
give a detailed account of how the concentration field is represented. 
Sec.~\ref{sec:IV} is dedicated  to a 
discussion of the results. The convection states are compared in
quantitative detail with simulations. We summarize our results in Sec.~\ref{sec:V}.

\section{System}
\label{sec:II}
We consider a convection cell of height $d$.
It contains a binary fluid of mean temperature $\bar{T}$ and mean
concentration $\bar{C}$ of the lighter component
confined between two perfectly heat conducting
and impervious plates. This setup is exposed to a vertical gravitational
acceleration $g$ and to a vertical temperature gradient $\Delta T/d$
directed from top to bottom.
The fluid has a density $\rho$ which varies due to temperature and
concentration variations governed by the linear thermal and solutal
expansion coefficients
$ \alpha = - \frac{1}{\rho}\frac{\partial\rho}{\partial \bar{T}} $ and 
$ \beta = - \frac{1}{\rho}\frac{\partial\rho}{\partial \bar{C}} $,
respectively. Its viscosity is $\nu$, the solutal diffusivity
is $D$, and the thermal diffusivity is $\kappa$. The thermodiffusion
coefficient $k_T$ quantifies the Soret coupling which describes the 
driving of concentration currents by temperature gradients.

The vertical thermal diffusion time $d^2/\kappa$ is used as
the time scale of the system and velocities are
scaled by $\kappa/d$. Temperatures are reduced by 
$\frac{\nu\kappa}{\alpha g d^3}$ and concentrations 
by $\frac{\nu\kappa}{\beta g d^3}$.
The scale for the pressure is given by $\frac{\rho\kappa^2}{d^2}$.
Then, the balance equations for mass, momentum, heat, and concentration
\cite{LL66,PL84} read in Oberbeck--Boussinesq approximation
\cite{HLL92}
\begin{eqnarray}
0 & = & - \Nabla \cdot \vec{u} \label{eq:baleqmass}\\
\partial_t\vec{u} & = & - (\vec{u} \cdot \Nabla) \vec{u}
 - \Nabla\left[p + \left(\frac{d^3}{\kappa^2}g\right) z\right]
+ \sigma \nabla^2\vec{u}
 + \sigma \left(T+C\right)\vec{e}_z \label{eq:baleqveloc}\\
\partial_tT & = & - \Nabla \cdot \vec{Q}\ =\ 
 - \Nabla \cdot \left[ \vec{u} T - \Nabla T\right]\label{eq:baleqheat}\\
\partial_tC & = & - \Nabla \cdot \vec{J}\ =\
 - \Nabla \cdot \left[ \vec{u} C - L\Nabla\left(C -\psi T\right) \right]\ .
 \label{eq:baleqconc}
\end{eqnarray}
Here, the currents of heat and concentration, $\vec{Q}$ and
$\vec{J}$ respectively, are introduced and $T$ and $C$ denote deviations of
the temperature and concentration fields, respectively, from their global
mean values $\bar{T}$ and $\bar{C}$. The Dufour effect \cite{HLL92}
that provides a coupling of concentration gradients into
the heat current $\vec{Q}$ and a change of the thermal diffusivity
is discarded in (\ref{eq:baleqheat}) since it is relevant
only in few binary gas mixtures \cite{LA96}.

Three parameters enter into the field equations
(\ref{eq:baleqmass})--(\ref{eq:baleqconc}): the Prandtl number
$\sigma=\nu/\kappa$, the Lewis number $L=D/\kappa$, and the separation
ratio $\psi=-\frac{\beta}{\alpha}\frac{k_T}{\bar{T}}$.
The latter characterizes the sign and the strength of the
Soret effect. Positive Soret coupling $\psi$ induces concentration gradients
parallel to temperature gradients. In this situation, the buoyancy
induced by solutal changes in density enhances the thermal buoyancy.
When the total buoyancy exceeds a threshold convection sets in, typically
in the form of squares for positive $\psi$. 

A fourth parameter, 
the Rayleigh number $R=\frac{\alpha g d^3}{\nu \kappa}\Delta T$
measuring the thermal driving of the fluid enters the description via the
boundary conditions of the temperature field (see below). 

The strength of the convection and its influence on convective
temperature transport can be measured by the Nusselt number 
$N = \frac{1}{R} \langle \vec{Q}\cdot\vec{e}_z \rangle_{xy}$ giving the ratio between the 
lateral average of the vertical heat current through the system and its conductive
contribution. In the basic state of quiescent heat conduction its value is thus $1$ 
and larger than $1$ in all convective states.

Solving the partial differential equations
(\ref{eq:baleqmass})--(\ref{eq:baleqconc}) requires boundary conditions for
the fields. We use realistic no slip conditions for the top
and bottom plates at $z=\pm \frac{1}{2}$,
$$ \vec{u}(x,y,z=\pm \frac{1}{2};t) = 0\ , $$
and assume perfectly heat conducting plates by 
$$ T(x,y,z=\pm \frac{1}{2};t) = \mp \frac{1}{2} R\ . $$
Furthermore, impermeability for the concentration is guaranteed by
\begin{equation}
\label{eq:Cbound}
\vec{e}_z\cdot\vec{J} =
 - L\partial_z\left(C-\psi T\right)(x,y,z=\pm \frac{1}{2};t) = 0\ .
\end{equation}

\section{Mode selection and Galerkin model}
\label{sec:III}

\subsection{Temperature and velocity field}
\label{sec:IIIA}
To describe the convective state we use the Galerkin method. We should stress that we
restrict ourselves to the description of extended patterns that are periodic in the lateral 
directions $x$ and $y$ with a certain lateral periodicity length $\lambda=2\pi/k$ and fixed
phases. We take $k=3.117$, i.e. the critical wave number of the pure fluid.

The temperature field which consists of a linear conductive profile
$-z$ and a convective deviation is truncated as
\begin{eqnarray}
\label{eq:Tansatz}
T(x,y,z;t) & = -R z & + T_{002}(t)\sqrt{2}\sin(2\pi z)\nonumber\\
& & + \left[T_{101}(t) \cos(kx) + T_{011}(t) \cos(ky)\right]\sqrt{2}\cos(\pi z)\\ \nonumber
& & + T_{112}(t) \cos(kx) \cos(ky)\sqrt{2}\sin(2\pi z)\ .
\end{eqnarray}
The indices $(l,m,n)$ of the amplitudes denote the expansion in $x$--, $y$-- and 
$z$--direction, respectively. If we restricted ourselves to 2D convective patterns only, e.~g.\ homogeneous in
$x$--direction, only the modes with the first index equal to zero would be
taken into account, reducing the temperature ansatz to that of a Lorenz model. 
In order to have an analogous representation of roll patterns in $y$--direction 
we also include the mode $T_{011}$. As we see in Eq.~(\ref{eq:Tansatz}) a further 
nonlinear mode, $T_{112}$, caused by the interaction of the two roll patterns is also 
introduced. The contribution of this mode vanishes for 2D patterns.
Note that only modes with an even index sum $l+m+n$ appear in the 
expansion, which is due to a mirror glide symmetry of the patterns studied \cite{H96}.

Choosing the first \emph{Chandrasekhar} function, ${\cal C}_1(z)$ \cite{HLBJ00,C81} as an 
approximation to the $z$--profile of the critical mode, the vertical component $w$ of the
velocity field reads
\begin{equation}
\label{eq:wansatz}
w(x,y,z,t) = \left[w_{101}(t) \cos(kx) + w_{011}(t) \cos(ky)\right] {\cal C}_1(z)\ .
\end{equation}
The $x$-- and $y$--components can be derived by using the incompressibility criterion 
(\ref{eq:baleqmass}). As the velocity field is dominated by its critical modes not only at 
onset but also far beyond, these two modes suffice to describe the patterns to be studied.
When discussing bifurcation diagrams we will use the amplitudes $w_{101}$ and $w_{011}$ as 
order parameters.

\subsection{Approximation of the boundary condition}
\label{sec:IIIB}
To select adequate concentration modes a detailed analysis of the
concentration balance and of the field structure of roll and square states
is necessary.

By introducing the combined field $\zeta = C-\psi T$ expanded as
\begin{equation}
\label{eq:zetaansatz}
\zeta(x,y,z,t) = \sum _{l=0}^\infty\sum _{m=0}^\infty\sum _{n=0}^\infty
\zeta _{lmn} \cos(lkx) \cos(mky) \textnormal{sc}_n(z)\ ,
\end{equation}
it is possible to fulfill the realistic boundary condition for the concentration 
current using appropriate trigonometric functions for the $\textnormal{sc}_n(z)$ to 
ensure $\partial_z \zeta = 0$ at the plates.

But as we aim to construct a few--mode model, we have to consider the balance
equations using the concentration field instead of the $\zeta$-field.
As discussed in \cite{HLM98,H96,W05} for stationary and travelling patterns in detail, the
reason lies in the balance equation for the $\zeta$-field:
\begin{equation}
\left(\partial_t+ \Nabla \cdot \vec{u} \right)\zeta = L\nabla^2 \zeta -\psi\nabla^2 T \ .
\label{eq:baleqzeta}
\end{equation}
The fluid mixtures which we refer to have Lewis numbers up to 2 orders of
magnitude smaller than the separation ratio. In this work we consider  $L =0.01$, $\psi
=0.1$ and $L =0.0045$, $\psi =0.23$ for example. Thus, in order to approximate
Eq.~(\ref{eq:baleqzeta}) consistently we have to take into account temperature modes which
are up to 2 orders of magnitude smaller than the $\zeta$-modes. Thus, despite the fact that
the temperature field is well described by the critical and the first nonlinear mode alone,
many more modes would be required when using Eq.~(\ref{eq:baleqzeta}). To avoid this
problem, we expand the concentration field itself. As a consequence, we can guarantee the 
boundary condition (\ref{eq:Cbound}) only approximatively.

\subsubsection{Approach}

\label{sec:IIIB1}
We adopt Hollinger's successful approximation for the boundary condition, which he proposed in  
\cite{H96}. His investigations on the concentration current induced by advection, 
diffusion and the Soret coupling support to demand impermeability only for the lateral
mean of the concentration field. In the lateral mean 
Eq.~(\ref{eq:Cbound}) becomes
\begin{equation}
\langle \vec{e}_z  \cdot \vec{J}\rangle_{xy} = -L\partial_z C_{lat}(z) -L R \psi N = 0 
\label{eq:Cboundconc_multi} \;\; .
\end{equation}
Here
\begin{equation}
C_{lat}(z) =\langle C(x,y,z;t)\rangle_{xy}
\end{equation}
denotes the lateral mean of the concentration field 
and $N$ is the Nusselt number introduced in Sec.~\ref{sec:II}.
This suggests the following multi--mode Galerkin ansatz for the concentration field
\begin{equation}
C(x,y,z;t) = - R \psi N z+ \sum _{l=0}^\infty\sum _{m=0}^\infty\sum _{n=0}^\infty
c _{lmn} \cos(lkx) \cos(mky) \textnormal{sc}_n(z) \ .
\label{eq:Cansatz_multi}
\end{equation}
In order to avoid introducing temperature modes in Eq.~(\ref{eq:Cansatz_multi}) via 
$N$ we approximate the boundary condition (\ref{eq:Cboundconc_multi}) by
$\partial_z \langle C(x,y,z;t)\rangle_{xy}(\pm 1/2)=-R \psi$. This deviates from the 
correct value by a factor equal to the Nusselt number $N=O(1)$.
This approximation can be understood as the leading term in an amplitude
expansion of $N$ around the conductive state with $N=1$. 

\subsubsection{Exact versus approximate concentration boundary condition}

In this subsection we present a brief comparison of the results of an
"exact" multi--mode Galerkin simulation \cite{HLBJ00} which 
fulfills the boundary conditions of the concentration field 
exactly by introducing the $\zeta$-field given by Eq.~(\ref{eq:zetaansatz}), with 
another multi--mode simulation. For the latter, we use the 
same field ansatz for temperature and velocity as for the exact one but the 
concentration field ansatz is given by Eq.~(\ref{eq:Cansatz_multi}), 
with $N$ replaced by 1. We refer to this simulation as using approximated boundary
conditions (ABCs) as opposed to exact boundary conditions (EBCs).

\begin{figure}[t]
\centerline{\psfig{figure=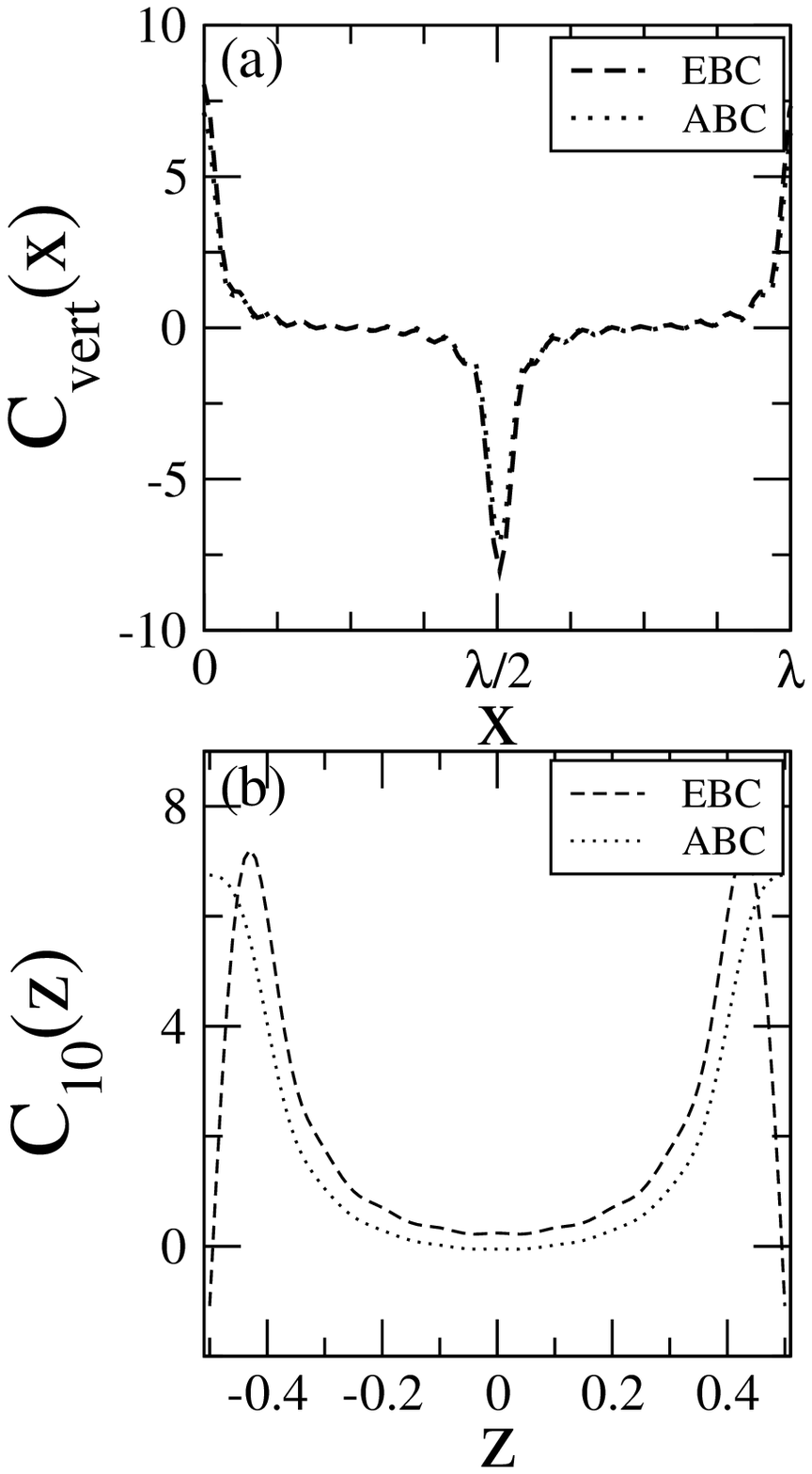,width=10cm}\vspace*{3mm}}
\caption{Comparison of multi--mode EBC simulations to multi--mode ABC 
simulations of roll convection. (a) Vertical mean $C_{\mathrm{vert}}(x)$ of the 
concentration field versus $x$. (b) First lateral Fourier mode $C_{10}(z)$ versus $z$. 
Parameters are $\psi =0.1, \sigma = 10, L=0.01$, and $r=1.5$.}
\label{fig:exact_approx}
\end{figure}

Fig.~\ref{fig:exact_approx}a shows the behavior of the vertical mean
$C_{\mathrm{vert}}(x)$ of the concentration for roll solutions
orientated in $y$--direction for a reduced Rayleigh number $r=R/R_c(\psi=0) = R/1708 =1.5$
where the homogenization of the concentration outside of narrow boundary layers is 
already apparent. Both, the EBC and the ABC simulation generate practically 
the same result; the ABCs have vanishing influence on the vertical mean.

The vertical dependence of the first lateral Fourier mode
\begin{equation}
C_{10}(z) = \frac{2}{\lambda^2}\int_0^\lambda\,dx\,dy\,C(x,y,z)\,\cos(kx)
\end{equation}
displayed in 
Fig.~\ref{fig:exact_approx}b shows also good agreement in the bulk. The influence of 
the different boundary conditions is restricted to the
boundary layer only. The ABC solution has a vanishing slope at the plates
as has any contribution to the concentration field except for the
vertical mean. The EBC solution ends with a finite slope at smaller
values. 

These results imply that the ABCs, that allow the construction of few--mode models, 
suffices to describe the concentration field in the bulk.

\subsection{Selecting the concentration field modes}

Analogous to the description of the temperature and the velocity field we want to include 
only few modes describing the concentration field. Still, we found that 10 modes are necessary to 
reproduce the bifurcation scenario. The concentration field ansatz which we finally used 
for the few--mode model is given by  
\begin{eqnarray}
\label{eq:Cansatz}
C(x,y,z;t)&=& -R \psi z + c_{001}(t)\sqrt{2}\sin(\pi z)+ c_{003}(t)\sqrt{2}\sin(3\pi z)\\\nonumber
&&+\left[c_{100}(t)\cos(kx)+c_{010}(t)\cos(ky)\right]\\\nonumber
&&+\left[c_{102}(t)\cos(kx)+c_{012}(t)\cos(ky)\right]\sqrt{2}\cos(2\pi z)\\\nonumber
&&+\left[c_{201}(t)\cos(2kx)+c_{021}(t)\cos(2ky)\right]\sqrt{2}\sin(\pi z)\\\nonumber
&&+\left[c_{300}(t)\cos(3kx)+c_{030}(t)\cos(3ky)\right]\ .
\end{eqnarray}
For all amplitudes the index sum $l+m+n$ is odd, reflecting the mirror glide symmetry already 
mentioned above. To check and justify the selection of these modes we analyze the lateral and 
vertical mean of the concentration field in the following subsection and thus demonstrate the relevance of these modes. 
We compare our results with the multi--mode simulations presented in \cite{HLBJ00}. 
Since we found similar results for both, square and roll solutions, we will 
concentrate on rolls here and do not present the detailed comparison of the square results.

\subsubsection{Lateral average of the concentration field}
\label{sec:IIIB2}
\begin{figure}
\centerline{\psfig{figure=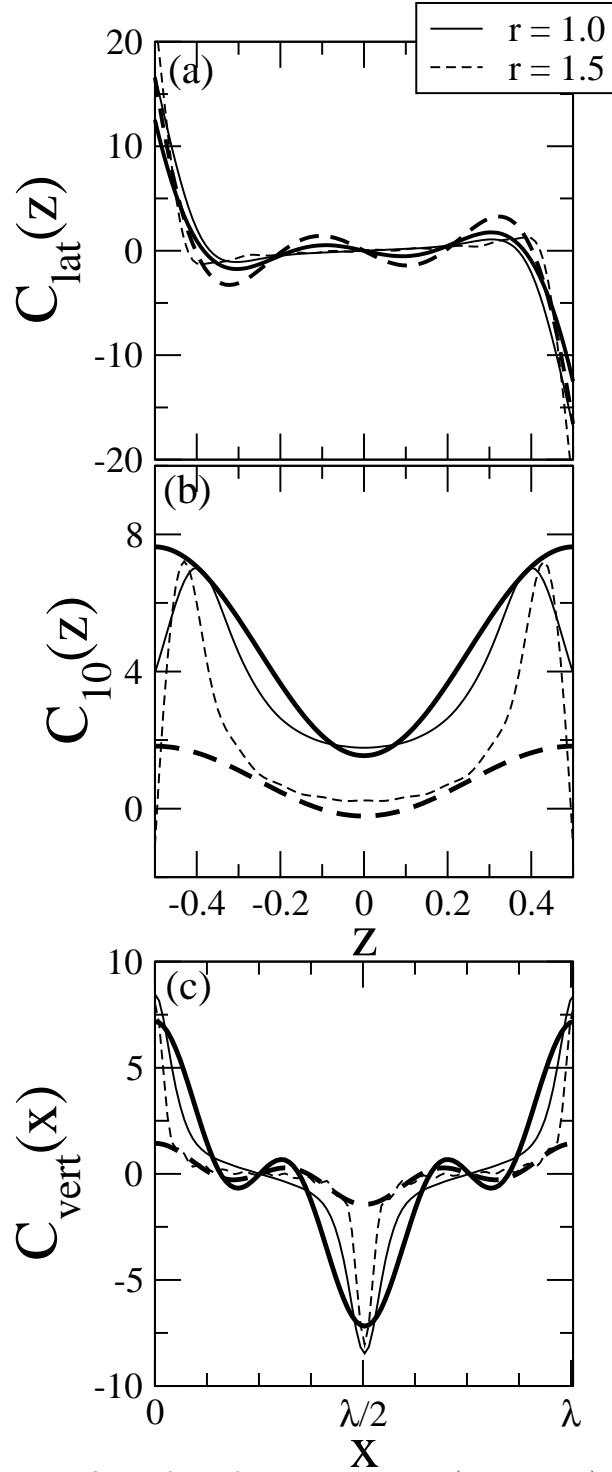,width=8cm}}
\caption{The concentration field of the few--mode model (thick lines) compared to EBC numerical
calculations (thin lines) at $r=1$ (solid) and $r=1.5$ (dashed) for rolls. (a) Lateral mean. 
(b) Lateral Fourier mode $C_{10}(z)$. (c) Vertical mean. Parameters are $\psi =0.1, \sigma = 10, L=0.01$.}
\label{fig:concentration_mean_field}
\end{figure}

The lateral mean $C_{\mathrm{lat}}(z)$ of the concentration field is determined by 
modes of the form $c_{00n}$. Our few--mode ansatz takes two such modes into account, namely $c_{001}$ and $c_{003}$. 
Figure~\ref{fig:concentration_mean_field}a shows $C_{\mathrm{lat}}(z)$ for two 
different reduced Rayleigh numbers $r$. For $r=1$ (thin solid
line), i.e., at the boundary between Soret and Rayleigh region, the homogenization of 
$C_{\mathrm{lat}}(z)$ in the bulk with pronounced concentration boundary
layers at the lower and upper plate becomes already apparent. The profile hardly 
changes when switching to a higher $r=1.5$ (thin dashed line). The two modes, $c_{001}$ and $c_{003}$, 
that are also taken into account describe this behavior well (thick lines), whereas one Fourier mode alone would 
be insufficient. 

\subsubsection{Critical modes of the concentration field}
\label{sec:IIIB3}
The critical modes have a lateral dependence $\cos {{kx}\choose{ky}}$ depending on 
the orientation of the rolls studied and are thus linear combinations of modes
with amplitude ${c_{10n}} \choose {c_{01n}}$. For each case, we again took two modes into 
account, ${c_{100}} \choose {c_{010}}$ and ${c_{102}} \choose {c_{012}}$. The vertical
variation of the first lateral Fourier mode $C_{10}(z)$ for rolls orientated in $y$--direction 
is shown in Fig.~\ref{fig:concentration_mean_field}b. For both, $r=1$ and $r=1.5$, the
field in the bulk is reproduced well, whereas the boundary layers are not properly resolved.
As we saw before, the reason for this does not lie in an insufficient number of modes but 
already in the ABC as the ABC multi--mode simulation has shown the same deficiency.

\subsubsection{Vertical average of the concentration}
\label{sec:IIIB4}
$C_{\mathrm{vert}}(x)$, the vertical mean of $C(x,y,z;t)$ is represented
by the first two lateral modes of the roll patterns, $\sin {{kx}\choose{ky}}$ 
and $\sin {{3kx}\choose{3ky}}$. In Fig.~\ref{fig:concentration_mean_field}c 
the vertical average for the multi--mode simulation is compared to the results of our 
few--mode model. The multi--mode result shows concentration peaks between the convection rolls 
and homogenization inside them. The inclusion of the higher harmonics in the few--mode
model allows to reflect this to some extent, at least for $r=1$ where the
concentration peaks are still relatively broad. The more narrow peaks at $r=1.5$ are 
not resolved anymore. However, even then there is still good agreement further away from the peaks.

\subsubsection{Further modes}
\label{sec:IIIB5}

The modes with lateral variation $\sin {{3kx}\choose{3ky}}$ are decoupled from the others
without the inclusion of further nonlinear modes. Adding the modes 
$c_{{201}\choose{021}}\propto \cos{{2kx}\choose{2ky}}\sin(\pi z)$ allows for driving
them via these new modes and $w_{{101}\choose{011}}$ in the nonlinearity of the 
concentration field equation (\ref{eq:baleqconc}). Studies of smaller models indicate that 
stable squares at onset are impossible without including these modes \cite{ML88,W05}.

Similarly, the modes $c_{{012}\choose{102}}$ discussed in Sec.~\ref{sec:IIIB3} serve to 
drive the mode $c_{003}$. While this mode would still be driven
by $w_{{101}\choose{011}}$ and $c_{{100}\choose{010}}$ due to the different 
$z$--expansion of $w$ and $C$, the coupling is weak compared to the one via
$c_{{012}\choose{102}}$.

\section{Results}
\label{sec:IV}
In this section, we elucidate the roll, square, and crossroll solutions of our model.
First, we discuss three different bifurcation scenarios. 
Then, in Sec.~\ref{sec:IVB}, we focus on the oscillating patterns generated in the few--mode
model in contrast to the oscillating crossrolls which appear in the exact simulations. 
Finally, in Sec.~\ref{sec:IVC}, we present the phase diagram of the few--mode model 
and discuss it in the light of numerical results.

\subsection{Bifurcation scenario}
\label{sec:IVA}
\begin{figure}[t]
\centerline{\psfig{figure=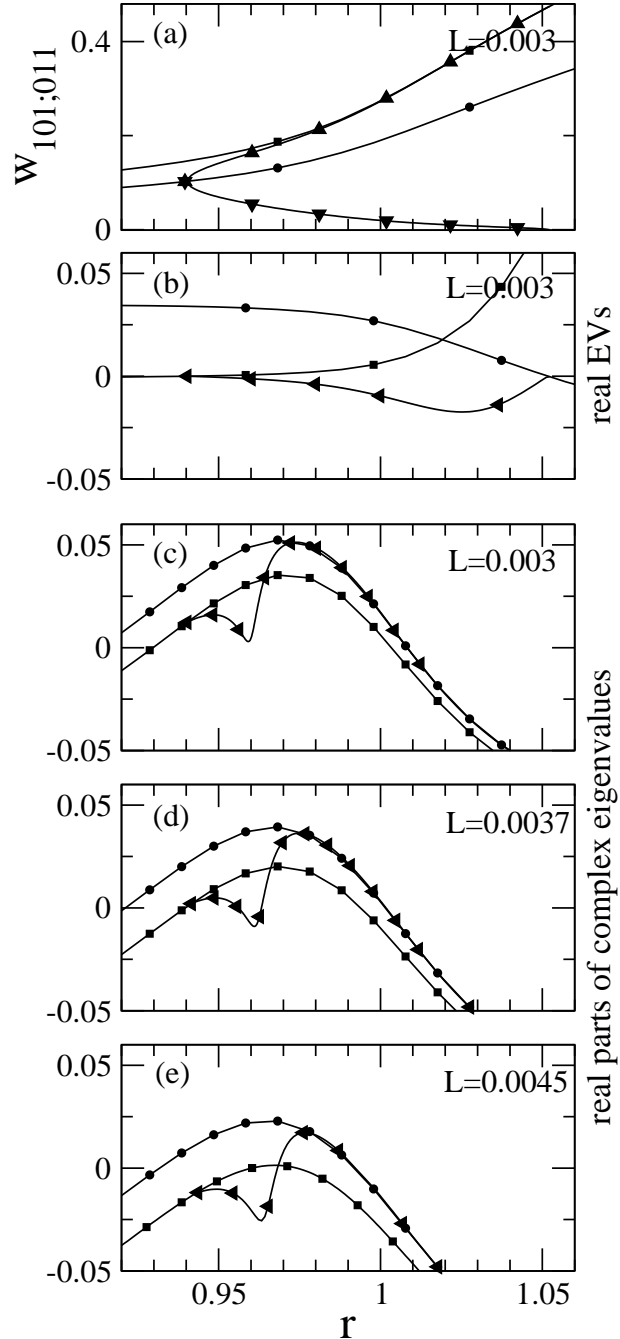,width=8cm}\vspace*{3mm}}
\caption{\small Bifurcation diagrams and corresponding eigenvalues of our 16 modes model. 
(a) bifurcation diagram of the velocity modes, (b) important real eigenvalues for (a). 
(c-e) real parts of the important complex eigenvalues. Square (roll) branches are 
symbolized by squares (circles). Thick lines and triangles mark the crossroll structure. 
Parameters are $\sigma=27$, $\psi=0.23$.}
\label{fig:bif_diagram_few}
\end{figure}

In Fig.~\ref{fig:bif_diagram_few}, results for the few--mode model are 
plotted for three different Lewis numbers in the interesting range of heating rates where 
the transition between small--amplitude stable squares in the Soret regime and higher--amplitude
stable rolls in the Rayleigh regime takes place. All plots presented were calculated
at $\psi = 0.23$ and $\sigma = 27$. These parameters can be realized by
ethanol--water mixtures and have also been studied in \cite{JHL98}.

The upper plot shows a bifurcation diagram representing the three different stationary
structures found. The square (S) branch is denoted by squares, the roll (R) branch by 
circles. Each is represented by a single curve: For square structures it is 
$w^S_{101} \equiv w^S_{011}$ while for rolls one of the amplitudes is zero, $w^R_{011}$ 
say, and thus only $w^R_{101}$ is plotted. 
It is $w^R_{101}/w^S_{101} \approx \sqrt{2}$ throughout the diagram, or in other words,
the sum $w^2_{101} + w^2_{011}$ is approximately the same for both structures. For the third
structure, the stationary crossrolls (CR), the two amplitudes $w^{CR}_{101}$ and 
$w^{CR}_{011}$ are different and nonzero, and are thus represented by two curves 
marked by up and down triangles respectively. The crossrolls bifurcate 
out of the square branch with equal amplitude but their difference grows with growing $r$ until
$w^{CR}_{101}$ meets the  roll branch and $w^{CR}_{011}$ becomes zero. The plot
has been calculated for $L=0.003$ but it does not change qualitatively and hardly quantitatively
for the values of $L$ considered below.

The second plot shows the real eigenvalues tied to the stationary instabilities for the same 
value of $L$. They are again identified by square, circle, and triangle symbols. On the  small $r$ side 
rolls exhibit a positive eigenvalue and are thus unstable. The square eigenvalue is negative even 
though this is hardly visible on the scale displayed; the smallness of the eigenvalue is a consequence 
of the very slow concentration dynamics ($L \ll 1$) in the Soret regime. On the large $r$ side the 
signs of the eigenvalues are reversed: rolls are now stable while squares are not. At the bifurcation 
points where the square eigenvalue crosses the zero axis the crossrolls appear and vanish again where the 
roll eigenvalue crosses the zero axis. In between, crossrolls are the only stationary stable 
structure and transfer this stability from the squares to the rolls. 

The stationary bifurcation properties of the few--mode Galerkin model, and in particular the 
sequence of stable squares, stable crossrolls, and stable rolls agree well with experimental and 
theoretical results for the system \cite{MS86,GPC85,JHL98} for not too small $L$. However, at very small 
$L$ another, time--dependent crossroll pattern appears in the vicinity of the bifurcation point from 
squares to stationary crossrolls. The time--dependent crossrolls emerge from an oscillatory 
perturbation and are thus represented by a complex eigenvalue.

The real parts of the most important complex eigenvalues for $L=0.003$ and two further Lewis numbers
in our few--mode model are presented in the lower three plots of Fig.~\ref{fig:bif_diagram_few}.
The results show that when $L$ is small enough, an oscillatory perturbation destroys the stability 
of the square pattern already before the bifurcation point where the stationary crossrolls appear. 
These crossrolls are then also oscillatory unstable and gain stability only at higher $r$ 
before meeting the roll branch (cases $L=0.003$ and $L=0.0037$). The real part of the crossroll 
eigenvalue has a local minimum at $r \approx 0.96$ such that crossrolls might temporarily gain 
stability before becoming unstable again. This happens for $L=0.0037$. They might also lose 
stability to an oscillatory perturbation only later on while being stable at the bifurcation point 
from the squares. This is the case for $L=0.0045$. For even larger $L$ the stationary crossrolls 
remain stable against oscillatory perturbations everywhere (not shown). Rolls can become unstable 
against oscillatory perturbations too, but we only found this to be the case at heating rates below 
the crossroll--roll bifurcation point where they are already stationary unstable.

\subsection{Oscillations}
\label{sec:IVB}

In \cite{JHL98}, Ch.~Jung et al. describe the behavior of oscillating crossrolls, 
as they appear in their multi--mode simulations at 
small $L$ between the regimes of stable squares and stable stationary crossrolls. 
In this structure, the leading amplitudes $w_{101}$ and $w_{011}$ oscillate 
around one of the square fixed points with opposite phase. With growing $r$ the 
oscillation becomes increasingly anharmonic until the oscillatory state disappears 
in a subharmonic bifurcation cascade, that is associated with an entrainment process.

\begin{figure}[t]
\centerline{\psfig{figure=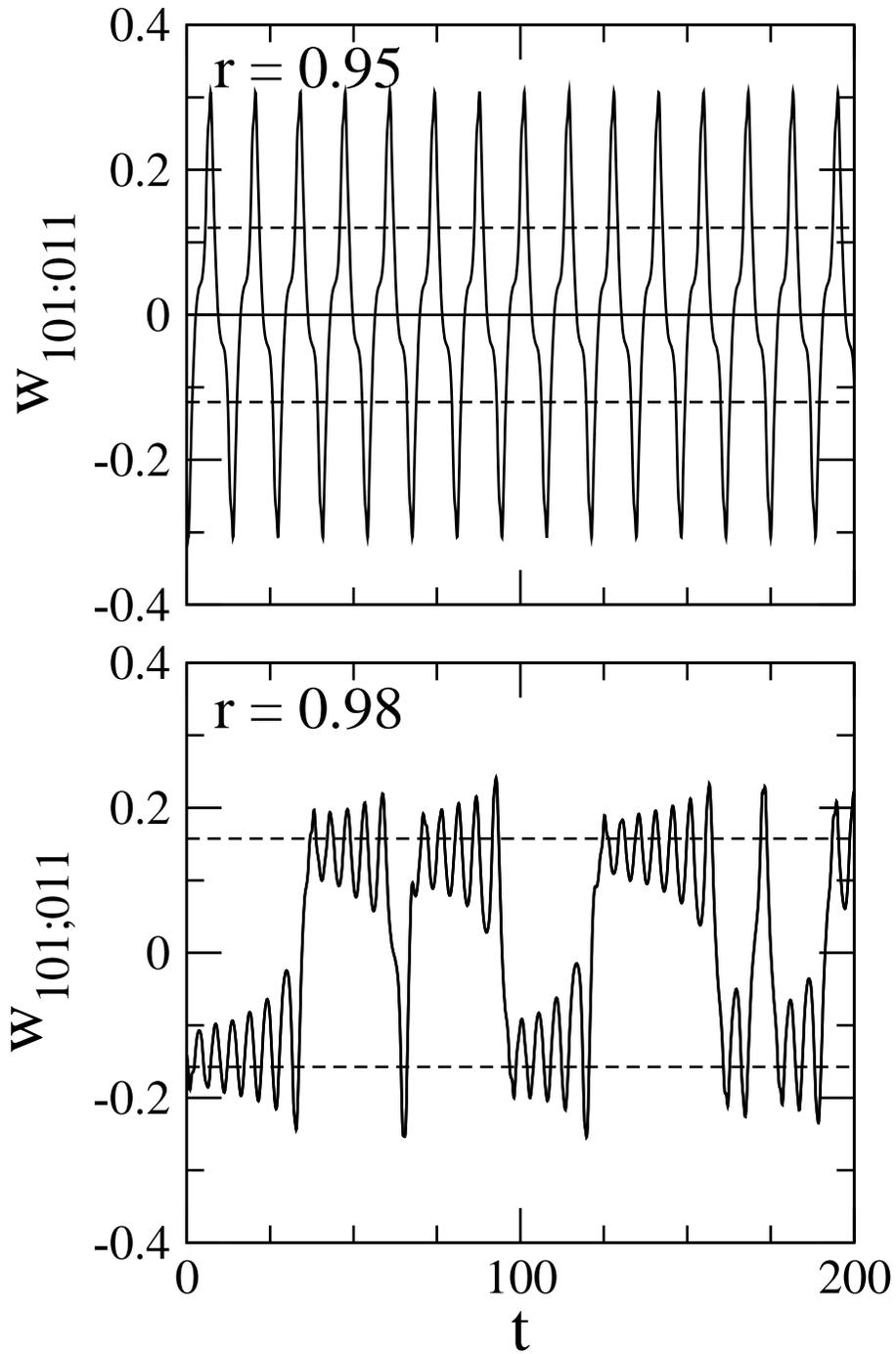,width=12cm,angle=270}\vspace*{3mm}}
\caption{Oscillations in the few--mode model. The dashed lines mark the fixed points of the 
stationary square patterns. Parameters are $\sigma=27$, $\psi=0.23$, and $L=0.0037$.}
\label{fig:osci_few}
\end{figure}

We studied time--dependent patterns in our few--mode model in the $r$--range where all three
stationary structures are unstable. We used the same time--integration method as in \cite{JHL98}. 
While the oscillatory instabilities of our few--mode model occur in the correct region of 
parameter space, the oscillating patterns of the model differ from the 
oscillating crossrolls discussed in \cite{JHL98}. We found different oscillatory regimes 
like those presented in Fig.~\ref{fig:osci_few} calculated for a Lewis number of $0.0037$. 

The pattern at $r=0.95$ is a 2D pattern in which only 
$w_{101}$ depends on time, while $w_{011}$ remains at zero (or vice versa). For $r=0.98$, 
$w_{101}$ and $w_{011}$ are equal, preserving the square
symmetry. This pattern oscillates around one of the square fixed points with growing amplitude,
followed by chaotic flips between the vicinities of the two square fixed points.

Since these time--dependent structures disagree qualitatively with those found
in the multi--mode simulations, we did not study them further. We conclude that a few--mode model
does not suffice to capture the nature of the oscillatory crossrolls.

\subsection{Phase diagram}
\label{sec:IVC}
\begin{figure}[t]
\centerline{\psfig{figure=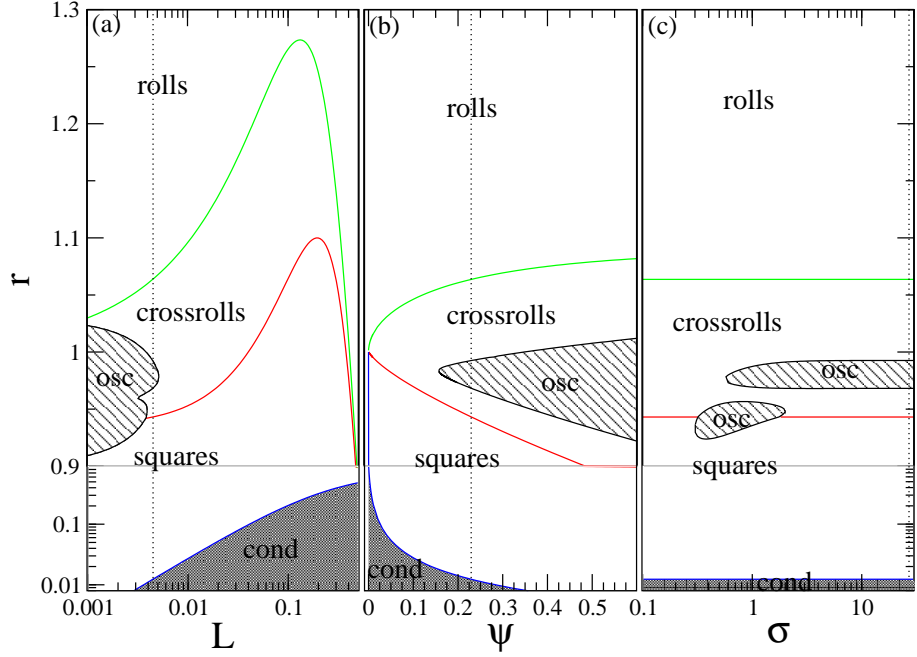,width=12cm,angle=270}\vspace*{3mm}}
\caption{Phase diagrams of stable convective states in the few--mode model. The dotted lines
refer to the parameters $L=0.0045$, $\psi =0.23$ and $\sigma = 27$, respectively. Two of them are kept 
fixed in each of the diagrams while the third is varied. The ordinate scale is logarithmic for $r< 0.8$ to
show the bifurcation threshold $r_{stat}(k=3.117;\psi,L)$ of the conductive state.}
\label{fig:PhaseI}
\end{figure}

\begin{figure}[t]
\centerline{\psfig{figure=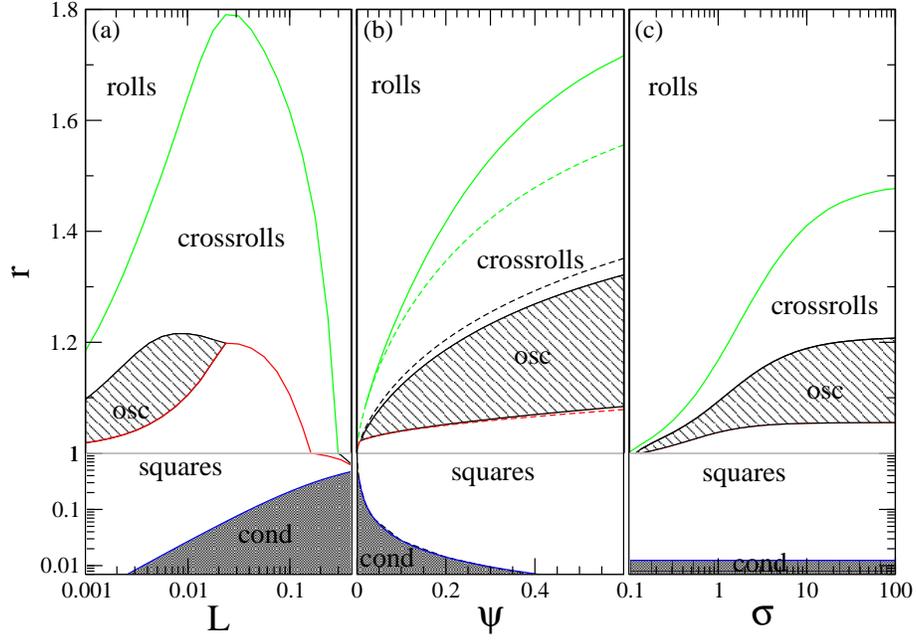,width=12cm,angle=270}\vspace*{3mm}}
\caption{Exact phase diagram of stable convective states in the EBC multi--mode simulation.
Parameters are chosen as in Fig.~\ref{fig:PhaseI}\,. In (b) we additionally illustrate
the phase diagram for ABCs (dashed lines).}
\label{fig:PhaseII}
\end{figure}

Figure \ref{fig:PhaseI} illustrates the phase diagrams of the few--mode model in 2D
planes of the 4D parameter space of $r$, $\psi$, $L$, and $\sigma$. 
In the following, the results in Fig.~\ref{fig:PhaseI} are compared 
to the EBC results (Fig.~\ref{fig:PhaseII}) taken from \cite{JHL98}.

We consider first the $L-r$ plane in Fig.~\ref{fig:PhaseI}a. Qualitatively, the phase
diagram is very similar to the full numerical EBC result.
For small enough $L$ the conductive state loses its stability against stable squares which in
turn lose stability at higher $r$ against oscillating structures discussed in the last 
subsection. The oscillations are then replaced by stationary 
crossrolls. Those finally merge with the roll branch. For $L > 0.005$ the oscillatory
regime is absent, and squares lose their stability directly to stationary crossrolls. At 
$L \approx 1$, outside of the plotted interval rolls are stable at onset. Note the dent 
in the region of oscillations at $L \approx 0.004$, $r \approx 0.96$ where the pattern sequence 
becomes more complicated. Three example cases have been studied in Sec.~\ref{sec:IVA}.

Fig.~\ref{fig:PhaseII}a looks qualitatively very similar. Only the dent in the oscillatory 
crossroll region is absent and the sequence of patterns with increasing $r$ is for small $L$ always 
squares -- oscillatory crossrolls -- stationary crossrolls -- rolls. However, there are two main
quantitative differences. First, the region of oscillations is larger in the multi--mode simulation; 
the direct squares -- stationary crossrolls transition happens only at $L>0.02$. The second 
important quantitative difference between multi--mode and few--mode model lies in the position of 
the phase space interface between roll and stationary crossroll patterns. The maximum in
Fig.~\ref{fig:PhaseI}a lies at $r=1.27$, whereas in Fig.~\ref{fig:PhaseII}a the maximum of
this curve lies at $r=1.79$\,. In \cite{JHL98}, the authors point out that the data
gained in numerical simulations become quantitatively unreliable for small $L$ combined with large
$r$. Comparing finite difference calculations and Galerkin models with different
spatial resolutions they found that the range of existence of crossrolls shrinks with decreasing
resolution which also explains the location of the crossroll--roll boundary in our
few--mode model.

Next, we want to study the $\psi$--dependence in Fig.~\ref{fig:PhaseI} and
Fig.~\ref{fig:PhaseII}\,. Note that our chosen fixed value of $L=0.0045$ is the one 
where the dent in the region of oscillations is located in Fig.~\ref{fig:PhaseI}a and where the 
above described sequence of patterns does not occur with growing $r$. Thus, the results in 
Fig.~\ref{fig:PhaseI}b look qualitatively different for this $L$ value from those of Fig.~\ref{fig:PhaseII}b. 
Stable stationary crossrolls exist above the oscillatory structures as well as below for all $\psi$ 
where oscillatory structures exist at all. This was already seen for the special case $\psi = 0.23$ which
is again marked by a dotted line. 

Finally, let us consider the $\sigma - r$ plane in Fig.~\ref{fig:PhaseI}c. The stationary stability 
curves are independent of $\sigma$. The reason lies in the ansatz for the velocity. Our ansatz 
(\ref{eq:wansatz}) takes only the critical modes into account, which eliminates the nonlinearity in 
Eq.~(\ref{eq:baleqmass}) and leads to $\sigma$--independent  stationary solutions. In the 
multi--mode diagram in Fig.~\ref{fig:PhaseII}, only the region of stable squares remains roughly 
independent of $\sigma$ whereas the range of existence of oscillatory and stationary crossrolls 
shrinks with decreasing $\sigma$. Again, there are qualitative differences to Fig.~\ref{fig:PhaseII}c
concerning the appearance of oscillations. Oscillatory regimes are found in two different regions 
of Fig.~\ref{fig:PhaseI}c and do not exist at small $\sigma$.
 
To summarize our comparison between few--mode model and multi--mode simulations, we found 
similar phase diagrams in the $L - r$ plane. However, there are qualitative differences 
for the $\psi - r$ and $\sigma - r$ plane
concerning the locations of oscillatory crossrolls due to the fact that the Lewis number 
$L = 0.0045$ studied in \cite{JHL98} lies within a small range of $L$--values where the 
pattern sequence is more complicated in the few--mode model.

Finally, we would like to point at another comparison between multi--mode simulations using 
ABCs and EBCs in Fig.~\ref{fig:PhaseII}b:
We plotted the results of the ABC multi--mode model in the $\psi-r$ plane (dashed lines). 
Compared to the EBC results the ranges of existence
for all occurring patterns are nearly the same. Only the phase space boundary between
rolls and crossrolls that sensitively hinges on the model size is shifted significantly. There
is no qualitative difference in the location of the oscillatory region as found in the 
few--mode model. These results underline again the usefulness of the ABCs especially considering
the fact that stability properties are in general more model--sensitive than fixed
point properties.

\section{Conclusion}
\label{sec:V}
This paper presents results of a new few--mode Galerkin model for Rayleigh--B\'enard convection 
in binary mixtures with positive separation ratio. With only 16 modes it is able 
to produce bifurcation and phase diagrams similar to those found in experiment or 
stemming from much more extensive simulations. In particular, it reproduces
a sequence of stable squares, crossrolls, and rolls. The success of 
this model is based on and is due to a carefully chosen concentration field ansatz. This ansatz
was obtained from comparisons with multi--mode simulations.

We used an approximation to the impermeability boundary condition allowing
us to keep the number of temperature field modes at a minimum. The concentration field 
ansatz then fulfills the impermeability condition at the plates in a strict sense only for the quiescent conductive 
state. Comparisons between results of multi--mode Galerkin simulations for rolls and squares
using exact and approximated boundary conditions show that both lead to essentially 
the same concentration field in the bulk. Significant differences appear only in the 
concentration boundary layers.
Further comparisons show that the ABC treatment also leads to qualitatively the same
phase diagram.  This is an interesting result as former theoretical models with permeable 
boundary conditions lead to very different features. It appears that the impermeability 
condition for the conductive state is the most important condition to fulfill. 

In the few--mode model, we found stable squares and unstable rolls bifurcating out of the 
quiescent conductive state for a wide range of parameters. A transition to stable rolls at higher 
temperature differences (Rayleigh regime) happens via a crossroll pattern. This pattern 
bifurcates forward out of the squares and merges with the rolls at larger $r$. Previous few--mode models also 
show an intermediate regime between Soret and Rayleigh regime, but the behavior of the system was 
different from the simulations.

The results for the phase diagrams prove the success of our model, as we 
have found qualitatively similar stability domains for the stationary structures. While our 
model fails to generate oscillating crossrolls as they appear in 
multi--mode simulations \cite{JHL98}, similar oscillatory structures have been found
in roughly the same parameter region.

Our model is analytically manageable, and further studies can possibly pave the 
way for a better understanding of the behavior and the role of the physically important modes.




\pagebreak

\end{document}